\documentclass[10pt, conference, compsocconf]{IEEEtran}
\usepackage{amsmath, graphicx}
\usepackage{xcolor}
\usepackage{caption}
\usepackage{subcaption}

\usepackage{amssymb}
\usepackage{pifont}

\newcommand{\Rmnum}[1]{\expandafter\@slowromancap\romannumeral #1@}
\usepackage{amsthm}
\newtheorem{definition}{Definition}
\newtheorem{theorem}{Theorem}
\newtheorem{exmp}{Example}[section]

\usepackage{multirow}
\newcommand{\subparagraph}{}
\usepackage{titlesec}
\usepackage{caption}
\usepackage{setspace}

\author{\IEEEauthorblockN{Chi-Ning Chou\textsuperscript{*, $\dagger$}, Yu-Jing Lin\textsuperscript{$\dagger$}, \\Ren Chen and Hsiu-Yao Chang}
\IEEEauthorblockA{
Department of Computer Science and \\
Electrical Engineering\\
National Taiwan University, Taiwan}
\and
\IEEEauthorblockN{I-Ping Tu}
\IEEEauthorblockA{Institute of Statistical Science\\
Academia Sinica\\
Taipei, Taiwan}
\and
\IEEEauthorblockN{Shih-wei Liao}
\IEEEauthorblockA{Department of Computer Science\\
National Taiwan University, Taiwan\\
liao@csie.ntu.edu.tw, corresponding author}
\thanks{* Currently a first-year Ph.D. student at Harvard University, USA.}
\thanks{$\dagger$ Students contributed equally to this work.}
}

\title{Personalized Difficulty Adjustment for Countering the Double-Spending Attack in Proof-of-Work Consensus Protocols}

\IEEEoverridecommandlockouts

\begin{document}

\maketitle


\begin{abstract}

Bitcoin is the first secure decentralized electronic currency system. However, it is known to be inefficient due to its proof-of-work (PoW) consensus algorithm and has the potential hazard of double spending. In this paper, we aim to reduce the probability of double spending by decreasing the probability of consecutive winning. We first formalize a PoW-based decentralized secure network model in order to present a quantitative analysis. 
Next, to resolve the risk of double spending, we propose the \emph{personalized difficulty adjustment (PDA)} mechanism which modifies the difficulty of each participant such that those who win more blocks in the past few rounds have a smaller probability to win in the next round. 
To analyze the performance of the PDA mechanism, we observe that the system can be modeled by a high-order Markov chain. Finally, we show that PDA effectively decreases the probability of consecutive winning and results in a more trustworthy PoW-based system.

\end{abstract}

\begin{IEEEkeywords}
personalized difficulty adjustment, proof-of-work, high-order Markov chain, state reduction
\end{IEEEkeywords}


\section{Introduction}
\label{sec:intro}

In 2008, Satoshi Nakamoto \cite{nakamoto2008bitcoin} published a breakthrough decentralized cryptosystem called Bitcoin, which ushered in an era of fully distributed trust network. Ten years have passed since its publication and Bitcoin is indeed a successful electronic currency. Many papers have been published for Bitcoin applications \cite{Ephraim2018}, however, very few are ever materialized. We believe that there are two main reasons for this phenomenon: the intrinsic overhead of the Bitcoin construction and the security hazards \cite{Harry2017Introduction} \cite{Sapirshtein2017} such as double spending.

The intrinsic overhead of Bitcoin lies in two parts of the system: the {\it information propagation} and the {\it proof-of-work mechanism}. The former is strongly related to the synchronization and consensus issue. 
Karame et al. \cite{karame2012two} \cite{karame2015misbehavior} initiated the study of fast payment in order to accelerate the transaction confirmation in Bitcoin and found out that the double spending probability is non-negligible. Bamert et al. \cite{bamert2013have} proposed {\it securing fast payments}, which improved the previous studies. They showed that under their construction the double spending probability diminishes to less than 0.088\%. Stathakopoulou et al. \cite{stathakopoulou2015faster} introduced a faster Bitcoin network based on the pipeline. They showed that increasing the locality of connectivity among each node can accelerate the information propagation. By implementing a {\it Content Distribution Network}, they achieved 60.6\% average speed up. From these works, we see that resolving the intrinsic overhead by accelerating the information propagation has limited performance.

As a result, we tried to fix the intrinsic overhead of Bitcoin network, or the PoW-based network, by proposing a new difficulty adjustment mechanism. The main focus is to strengthen the power of PoW so that it can guarantee stronger security.

\subsection{Proof-of-Work (PoW)}
In a PoW system, every participant has the right to be the verifier, but with a different probability. The probability to be selected as a verifier depends on how much he or she has devoted to the system. That is, the more one contributes to the system, the higher probability he or she has to be selected as a verifier. The mechanism to evaluate the amount of devotion is the so-called {\it proof-of-work}. Practically speaking, each participant keeps computing hash values of the header of a block. And the first one who finds a small enough hash value will be regarded as the verifier for the current block. 

\subsection{Double Spending}
However, there is a potential hazard that the verifier is an attacker. Attackers might modify the transaction record and benefit themselves. In \cite{nakamoto2008bitcoin}, Satoshi referred to this kind of situation as double spending in which the verifier spends the same money twice, which is definitely not allowed to happen in a trusted system. 

The original double spending scenario in \cite{nakamoto2008bitcoin} considers the possibility that some attackers in Bitcoin system build their own blockchain instead of mining on the main chain. 
The attackers build a private blockchain that is longer than the public chain. 
Thus, other participants in the system will adopt the longer one, which in this case is the attackers' private blockchain. 
The attackers produce some fake transactions in its private blockchain, usually by removing the records of their spendings so as to spend that money again in the future.
Once these fake transactions such as a double spending one, merge into the main blockchain, the trust of the system will break down. 

\subsection{High-order Markov Chain}

Raftery \cite{raftery1985model} proposed the Mixture Transition Distribution model (MTD) in 1985 which models the high-order Markov chain with a lag coefficient. Later, Berchtold and Raftery \cite{berchtold2002mixture} generalized the idea into multi-matrix MTD, infinite-lag MTD, spatial MTD, etc., which can be used in various applications. In 2005, Ching et.al. \cite{ching2005computation} \cite{ng2006markov} relaxed the constraints in MTD model and yielded a more general results. Recently, Li and Ng \cite{li2014limiting} used probability tensor to model the high-order Markov chain, and found some sufficient conditions for the existence and uniqueness of stationary distribution.

However, these previous works are not quite the same as what we focus here. They generalized the Markov property into high-order Markov chain, which conditions on more than one past states, and emphasized on the stationary distribution or asymptotic behaviors of a single state. Here, we not only utilize the high-order Markov property but also obtain the stationary distribution or asymptotic behaviors of a sequence of states. In other words, the event we care about is a period of time or state, not a single snapshot. 

In personalized difficulty adjustment PoW system, what we concern is the double spending events in the scenario of consecutive winning by the same party. This paper presents the model to minimize the likelihood of spending the same coin twice and its corresponding results in Gcoin \cite{Tseng2018}\cite{Liao2014}.


\section{Model}\label{sec:model}
We use a general stochastic process to model the PoW system and construct the personalized difficulty adjustment PoW system step by step. We start from the traditional PoW system in Section~\ref{sec:traditionalbitcoin}, then we introduce the concept of personalized difficulty adjustment in Section~\ref{sec:dynamicdifficultybitcoin} and give an example of difficulty function in Section~\ref{sec:difficultyfunction}.

\subsection{Traditional PoW system}\label{sec:traditionalbitcoin}
First, note that the system is basically a discrete system. Namely, the system is composed of a sequence of blocks, and it's sufficient for us to use the index of each box to order the configurations in the system. In the following construction, we use the small letter $b$ to denote the order index of the block we are at.

Next, there are $n$ participants in the system competing for solving hash values. Each of them intrinsically has his own computing power, denoted as $C_i{(b)}$, where $i$ refers to the index of the participant and $b$ is the block index. Besides the computing power, there is a time-varying parameter recording the difficulty of the system. We denote the difficulty for competing block $b$ as $D(b)$, with computing power $\{C_i(b)\}$ and difficulty $\{D(b)\}$. It is sufficient to model the traditional PoW system with a two-tuple stochastic process as follows.
\begin{definition}[traditional PoW system]
	\mbox{}
    
	A traditional PoW system is a 2-tuple 
	$$\mathcal{B}_{trad.} = (\{C_i(b):i=1,...,n;\ b\in\mathbb{N} \}, \{D(b):b\in\mathbb{N} \}),$$
	where $C_i(b)$ is the computing power of player $i$ and $D(b)$ is the difficulty of the system at the $b^{th}$ block respectively.
\end{definition}

At block $b$, participant $i$ has computing power $C_i(b)$ and is assigned with difficulty $D(b)$. Each of the participants then keeps computing hash function until one of them finds a valid hash value. The probability of participant $i$ to win at block $b$ is proportional to $\frac{C_i(b)}{D(b)}\sim C_i(b)$. This observation is formally stated in Theorem~\ref{thm:proportional} and is proved in Appendix~\ref{sec:proofthm1}. 

\subsection{Personalized Difficulty Adjustment (PDA) PoW System}\label{sec:dynamicdifficultybitcoin}
\begin{figure}[b]
	\captionsetup{name=Figure}
	\centering
	\includegraphics[scale=0.45]{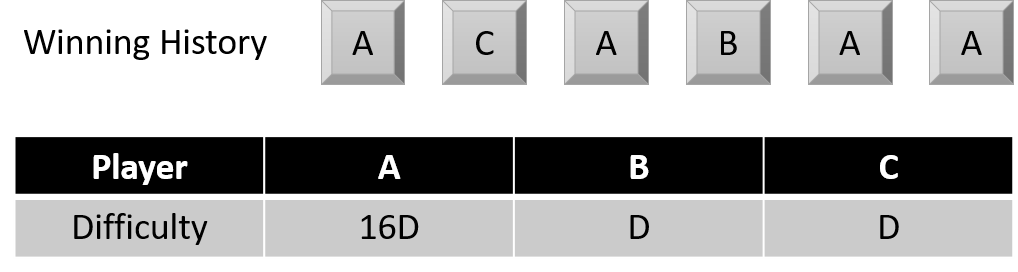}
	\caption{Example of personalized difficulty adjustment.}
	\label{fig:exampleofdynamicdifficulty}
\end{figure}

To simplify the problem, we assume all the nodes in PoW networks have the same computing power. A PDA PoW system modifies the dynamic parameter of itself according to the recent outcomes in the system. For example, in 
this system [Figure ~\ref{fig:exampleofdynamicdifficulty}], the difficulty index of each player will be configured according to the past winners in recent blocks. Suppose player A wins 4 blocks in the past 6 blocks, then his difficulty will be higher than player B who only wins 1 block in that period. Intuitively, we can think of the scenario as a dynamic stochastic process in which the transition mechanism will depend on a period of past results.

To formalize a PDA PoW system, there are three elements we need to add into the system: the winning history, the number of past blocks we concern, and the difficulty function.

The winning history records who wins the block from the beginning till current block $b$ denoted as $\{W{(b)}\}$, where $W{(b)}$ denotes the winner at the $b$th block. Next, we denote the number of past history as $k$. That is to say, if we are at block $b$, then we are going to consider the winner at blocks $b-1,b-2,...,b-k$. The difficulty function is specified by $\Psi_k$ from the past winning history to a difficulty assignment for every player. Thus, we define PDA PoW system as follows:

\begin{definition}[personalized difficulty adjustment proof-of-work system]
	\mbox{}
	
	A PDA PoW system with $n$ participants (players) is a 5-tuple 
	\begin{align*}
    \mathcal{B}=( & \{C_i{(b)}:i=1,...,n;\ b\in\mathbb{N} \}, \\ 
    & \{D_i{(b)}:i=1,...,n;\ b\in\mathbb{N}\}, \\ 
    & \{W{(b)}: b\in\mathbb{N}\},k,\Psi_k ),
    \end{align*}
	where
	
	\begin{itemize}
		\renewcommand\labelitemi{--}
		\item $C_i{(b)}\in\mathbb{C}$ denotes the computing power of player $i$ at block $b$. For simplicity, if not specified, we assume the computing power of each player is a constant and remains all the same. Intuitively, we can think of computing power as internal power.
		\item $D_i{(b)}\in\mathbb{D}$ denotes the difficulty of player $i$ at block $b$. Intuitively, we can think of difficulty as external power.
		\item $W{(b)}\in\{0,1,...,n-1\}$ denotes who wins at block $b$.
		\item $k$ is the number of blocks we look into the past.
		\item $\Psi_k$ is the difficulty function that looks into $k$ past blocks.
	\end{itemize}
\end{definition}

The process is actually overdetermined. The difficulty $\{D_i{(b)}\}$ are recursively defined by the winning history $\{W(b) \}$, the number of blocks we look in the past $k$ and the difficulty function $\Psi_k$. Similarly, we denote the probability of the $i$th player to compute for block $b$ as $P_i(b)$. And for convenience, we denote the recent $m$ histories $\{W^{(b)}, W^{(b-1)}, ..., W^{(b-m+1)} \}$ as $m$-history.

The probability of winning matters the most. With some probability argument, we can derive a simple relation between winning probability, difficulty, and computing power stated in the following theorem.
\begin{theorem}\label{thm:proportional}
	The winning probability of player $i$ at block $b$ is proportional to the winning probability divided by the difficulty. That is,  $P_i^{(b)}\sim \frac{C_i{(b)}}{D_i{(b)}}$. The proof is left in Appendix~\ref{sec:proofthm1}.
\end{theorem}
Intuitively, Theorem~\ref{thm:proportional} tells us that the winning probability of each player is proportional to the ratio of computing power and the assigned difficulty. As a result, once this ratio is approximately the same for all participants, then the winning probability will close to uniform.

With Theorem~\ref{thm:proportional}, we calculate the winning probability for player $i$ at block $b$ by dividing its ratio $\frac{C_i{(b)}}{D_i{(b)}}$ with the summation of all ratio $\sum_{j=0}^{n-1}\frac{C_j{(b)}}{D_j{(b)}}$. 
And Theorem~\ref{thm:proportional} tells us the winning probability is proportional to the ratio of computing power and difficulty. Thus, it's convenient for us to analyze the whole system.

\subsection{Difficulty function}\label{sec:difficultyfunction}
\begin{definition}[difficulty function]
	\mbox{}
	
	A difficulty function that considers $k$ past blocks $\Psi_k:\{0,1,...,n-1\}^k\rightarrow [0,1]^n$ is a mapping from $k$ past history $W{(b-1)}, W{(b-2)},..., W{(b-k)}$ to a difficulty vector for $n$ players: $D_1{(b)}, D_2{(b)},...,D_n{(b)}$.
	
	Also, with Theorem~\ref{thm:proportional}, we assume $\sum_{j=0}^{n-1}D_j{(b)}=1$ for simplicity.
\end{definition}

\begin{exmp}[$\alpha$-exponential non-ordered difficulty function]
	Denote the winning times of  player $i$ in the past $k$ blocks from current block $b$ as $w_i{(b)}=\sum_{j=1}^k\mathbf{1}_{\{W{(b-j)}=i\}}$. An $\alpha$-exponential non-ordered difficulty function $\Psi_k^{\alpha}$ maps
	\begin{align*}
	D_i{(b)}&=\big\langle\Psi_k(W{(b-1)}, W{(b-2)},...,W{(b-k)})\big\rangle_i\\
	&=\frac{\alpha^{w_i{(b)}}}{\sum_{j=0}^{n-1}\alpha^{w_j{(b)}}}
	\end{align*}
	
\end{exmp}
Suppose we take $\alpha = 2,n=3, k=3$ and every player has the same computing power. If the 3-history right now is $(W{(b)}=0,W{(b-1)}=2,W{(b-2)}=0)$, then the winning times of each player is:
\begin{align*}
w_0{(b+1)}=2, w_1{(b+1)}=0, w_2{(b+1)}=1
\end{align*}
The difficulty for each player at block $b+1$ is:
\begin{align*}
D_0{(b+1)} = \frac{2^{w_0{(b+1)}}}{\sum_{j=0}^2 2^{w_j{(b+1)}}} = \frac{2^2}{2^2+2^0+2^1} = \frac{4}{7}\\
D_1{(b+1)} = \frac{2^{w_1{(b+1)}}}{\sum_{j=0}^2 2^{w_j{(b+1)}}} = \frac{2^0}{2^2+2^0+2^1} = \frac{1}{7}\\
D_2{(b+1)} = \frac{2^{w_2{(b+1)}}}{\sum_{j=0}^2 2^{w_j{(b+1)}}} = \frac{2^1}{2^2+2^0+2^1} = \frac{2}{7}\\
\end{align*}
The winning probability for each player at block $b+1$ is: 
\begin{align*}
Pr[0 | 0, 2, 0] = \frac{1/D_0{(b+1)}}{\sum_{j=0}^2 1/D_j{(b+1)}} = \frac{1}{7}\\
Pr[1 | 0, 2, 0] = \frac{1/D_1{(b+1)}}{\sum_{j=0}^2 1/D_j{(b+1)}} = \frac{4}{7}\\
Pr[2 | 0, 2, 0] = \frac{1/D_2{(b+1)}}{\sum_{j=0}^2 1/D_j{(b+1)}} = \frac{2}{7}
\end{align*}
p.s. we denote $Pr[W{(b+1)}=x_1|W{(b)}=x_2,W{(b-1)}=x_3,W{(b-2)}=x_4]$ as $Pr[x_1|x_2,x_3,x_4]$

This example shows the intuition of $\alpha-$exponential non-ordered difficulty function: the winning probability at block $b+1$ is proportional to $\alpha^{w_i{(b+1)}}$.

\section{Rate of consecutive winning}\label{sec:rateofconsecutivewinning}
Now we want to examine how well the PDA mechanism prevents us from consecutive winning.

We define the scenario we care about.
\begin{itemize}
	\item {\bf Setting}: A PDA PoW system with $n$ players: $\mathcal{B}=(\{C_i{(b)}:i=1,...,n;\ b\in\mathbb{N} \}, \{D_i{(b)}:i=1,...,n;\ b\in\mathbb{N}\}, \{W{(b)}: b\in\mathbb{N}\},k,\Psi_k )$. Assume the  computing power is constant for each player, i.e. $C_i^{(b)}=C_j^{(b')},\forall i,j\in\{0,...,n-1\},\ \forall b,b'$.
	\item {\bf Goal}: We want to know the probability of player 1 consecutively winning for $k$ time, i.e. $\lim_{n\rightarrow\infty} Pr[W(n)=1,W({n-1})=1,...,W({n-k+1})=1]$.
\end{itemize}

First, we consider the case with no difficulty function (or we choose the $1-$exponential non-ordered difficulty function). 

\subsection{No difficulty function}
As the winning probability for each player at a single block is all the same: $\frac{1}{n}$, the probability of player $1$ to consecutively win $k$ blocks is $(\frac{1}{n})^k$. From another point of view, we can think of this as a no-difficulty-case as the winning probability of each block is i.i.d while, in other cases the winning probability of each block is correlated.

\subsection{Arbitrary difficulty function}
And then we consider choosing an arbitrary difficulty function $\Psi_k$. However, since the winning probability of each block is correlated, we can not simply utilize the i.i.d. property to calculate the goal: $\lim_{n\rightarrow\infty} Pr[W(n)=1,W({n-1})=1,...,W({n-k+1})=1]$. Instead, joint probability of consecutive blocks should be considered as:
\begin{align*}
&Pr[W{(b+1)}=i|W{(b)},...,W{(1)}]\\
&=Pr[W{(b+1)}=i|W{(b)},...,W{(b-k+1)}]
\end{align*}

We find out that system obeys {\bf high-order Markov chain}, or {\bf k-th order Markov property}.

\begin{definition}[high-order Markov chain]
The idea of high-order Markov chain is as follows.

The transition probability of a stochastic process is only conditioned on the previous $k$ events. Formally, suppose $(X_n)$ is a $k$-th order Markov chain over state space $\mathcal{X}$, then for $n\geq k$ and $\forall x_i\in\mathcal{X},0\leq i\leq n-1$,
\begin{align*}
& P[X_n=x_0 | X_{n-1} = x_1,...X_1 = x_{n-1}] \\
& = P[X_n=x_0 | X_{n-1}=x_1,...,X_{n-k}=x_k]
\end{align*}
\end{definition}

Therefore, we can directly encode the winning probability into a transition matrix as we regard each possible $k$ history as a state. Formally speaking, we define the state space of the Markov chain as $\mathcal{M}(\mathcal{B})=(S,P)$, where $S$ is the state space and $P$ is the transition probability function.

\begin{itemize}
	\renewcommand\labelitemi{--}
	\item $S=\{1,...,n\}^k$: the $k$ history.
	\item $P:S\times S\rightarrow[0,1]$: suppose $s=(s_1,...,s_k), s'=(s'_1,...,s'_k)$ then
	
    \begingroup
	\begin{align*}
	P(s,s')= \left\{ \,
	\begin{IEEEeqnarraybox}[][c]{l?s}
	\frac{\frac{C_{s'_1}{(b)}}{\big\langle \Psi_k(s) \big\rangle_{s'_1}}}{\sum_{i=0}^{n-1}\frac{C_{i}{(b)}}{\big\langle \Psi_k(s) \big\rangle_{i}}} , \\ if \ s_i=s'_{i+1}\forall i=1,...,k-1, \\ \\
	0 , otherwise. \\
	\end{IEEEeqnarraybox}
	\right.
	\end{align*}
    \endgroup
	where $\big\langle \Psi_k(s) \big\rangle_{i}$ is the difficulty for player $i$ conditioned on history $s$.
\end{itemize}

Let $\mathbb{W}_k{(b)}$ be the probability vector over $S$ denoting the probability of $k$ past history started from block $b$. Then, the stationary distribution of $k$ past history is $\hat{\mathbb{W}_k}$ that satisfies $\hat{\mathbb{W}_k}=\lim_{b\rightarrow\infty}\mathbb{W}_k{(b)}$. Or, equivalently, $\hat{\mathbb{W}_k}=P\hat{\mathbb{W}_k}$.

\subsection{$\alpha-$exponential non-ordered difficulty function}
Suppose we choose the $\alpha-$exponential non-ordered difficulty function in a $n$ participants, $k$ past history PDA PoW system $\mathcal{B}$. Then, we get $\mathcal{M}(\mathcal{B})=(S,P)$, where $S=\{1,...,n\}^k$ and 

\begingroup
\begin{align*}
P(s,s')= \left\{ \,
\begin{IEEEeqnarraybox}[][c]{l?s}
\frac{1/\frac{\alpha^{w_{s'_1}}}{\sum_{j=0}^{n}\alpha^{w_{j}}}}{\sum_{i=0}^{n-1}1/\frac{\alpha^{w_{i}}}{\sum_{j=0}^{n}\alpha^{w_{j}}}} = \frac{1/\alpha^{w_{s'_1}}}{\sum_{i=0}^{n-1}1/\alpha^{w_i}} , \\ if \ s_i=s'_{i+1}\forall i=1,...,k-1, \\ \\
0 , otherwise. \\
\end{IEEEeqnarraybox}
\right.
\end{align*}
\endgroup





Here we calculate the probability of consecutive winning with the different number of past history and different difficulty functions. The number of participants is 5.

\begin{table}[h]
	\centering
	\caption{Probability of consecutive winning. $n=5$.}
	\resizebox{\columnwidth}{!}{\begin{tabular}{|l|l|l|l|l|}
        \hline
		                          & 2       & 3       & 4       & 5        \\ \hline
		No difficulty             & 4e-2    & 8e-3    & 1.6e-3  & 3.2e-4   \\ \hline
		2-exponential non-ordered & 2.34e-2 & 1.59e-3 & 6.14e-5 & 1.35e-6  \\ \hline
		5-exponential non-ordered & 1.12e-2 & 1.64e-4 & 6.38e-7 & 6.74e-10 \\ \hline
	\end{tabular}}
	\label{table:consecutivewinning}
\end{table}
\vspace{-0.5em}

\subsection{Calculation issue}
We have discussed the probability of consecutive winning with different difficulty functions. And in Table~\ref{table:consecutivewinning} we show that the probability of consecutive winning drops down obviously as we increase the number of past history and the difficulty ratio $\alpha$.

However, once we increase the number of participants, the size of transition matrix drastically increases. For example, if we consider 4 past history, the number of states in the transition matrix of 5 participants is $5^4 = 625$. As we consider 20 participants, the number will become $20^4=160000$, which hardly can be computed by a normal computer! As a result, once we create a new difficulty function and want to see its performance of preventing from consecutive winning, we cannot efficiently compute the results if there is a large number of participants with the above calculation model.

When looking deeper into the transition matrix, we can find out that there are so many 0's. Namely, the matrix is extremely sparse. 

\section{State reduction}\label{sec:reduction}

While the number of states grows up, the outcomes are full of {\bf super-symmetry}. 
We intuitively put them all together and reduce the number of outcomes, which will drastically reduce the computation.

\subsection{Abstract model}
We find out that PDA PoW system and the basic dynamics share many similarities. However, some parameters in PDA PoW system change over time according to the outcome makes it more complicated. Therefore, we divide the system into two parts: The base rules and the parameters. Moreover, the parameters can also be categorized into fixed parameters and dynamic parameters. 

	
The PDA PoW system has the base rules analogous to the basic mechanisms such as mining policies, proof of work, timestamps, etc. The parameters are the number of players, computing power, difficulty function, difficulty etc. 
We formalize the abstract model of PDA PoW system as follows:


\begin{table}[h]
\centering
\caption{Analogy in PDA PoW system.}
\label{table:analogy}
\def\arraystretch{1.5} 
\resizebox{\columnwidth}{!}{\begin{tabular}{|c|l|l|}
\hline
\multicolumn{2}{|l|}{Basic rules}                    & Mining policies, proof of work, timestamps. \\ \hline
\multirow{3}{*}{Parameters} & \multirow{2}{*}{Fixed} & Difficulty function, number of players*,    \\
                            &                        & computing power.                            \\ \cline{2-3} 
                            & Dynamic                & Difficulty.                                 \\ \hline
\end{tabular}}
\end{table}
\vspace{-0.5em}

We define the fixed and dynamic parameters respectively as follows:
\begin{definition}[fixed parameters]
	\mbox{}
	
	The fixed parameters of PDA PoW system, denoted as $\theta_f$, is a 4-tuple
	$$\theta_f:=\{n,\{C_i\},k,\Psi_k \}$$
	, where
	\begin{itemize}
		\item $n$: number of players.
		\item $\{C_i\}$: computing power of each player.
		\item $k$: number of history considered in the difficulty function.
		\item $\Psi_k$: the personalized difficulty function.
	\end{itemize}
\end{definition}

 \begin{definition}[dynamic parameters]
	\mbox{}
	
	The dynamic parameters of PDA PoW system on block $b$, denoted as $\theta_d^{(b)}$, is a 2-tuple
	$$\theta_d^{(b)}:=\{\{D_i^{(b)}\}, \{W^{(b)}\} \},$$
	 where
	\begin{itemize}
		\item $\{D_i^{(b)}\}$: the difficulty of each player of block $b$.
		\item $\{W^{(b)}\}$: the winning history in the system. 
	\end{itemize}
\end{definition}



\subsection{Framework}
There are three steps in the reduction process:
\begin{enumerate}
	\item Reduce states.
	\item Transition matrix.
	\item Stationary distribution.
\end{enumerate}
In the first step, {\bf Reduce states}, we scan through all possible past configurations and generate reduced states. Next, {\bf Transition matrix}, we construct the corresponding transition matrix according to the reduced state, the basic parameters of the system, and the decay parameter of the exponential non-ordered model. Finally, {\bf Stationary distribution}, we use an iterative method to find the stationary distribution of reduced transition matrix and obtain the stationary probability of consecutive winning.

\subsubsection{Reduce states}
We construct a mapping from the standard state space to the reduced state space based on the intuition in Section~\ref{sec:reduction}. Formally, we define the standard state space and reduced state space as follows:

\begin{definition}[standard state space]
	The standard state space of the PDA PoW system, denoted as $S_s$, is the $m$-Cartesian product over the player space.
	$$S_s:=\{0,1,...,n-1\}^m,$$
	 which is the state space for $m$-history.
\end{definition}

\begin{definition}[reduced state space]
	The reduced state space of the PDA PoW system, denoted as $S_r$, is a subset of $S_s$ defined as follow:
    \begingroup
	\begin{align*}
	& S_r:=\{(s_1,...,s_m)\in S_s: \\
    & \left\{
    \begin{array}{ll}
	\forall 0<i<j<m, \sum_{k=1}^{m}\mathbf{1}_{\{s_k=i\}} \geq \sum_{k=1}^{m}\mathbf{1}_{\{s_k=j\}} \\
	\forall 0<i<j<m, \mbox{if}\ \sum_{k=1}^{m}\mathbf{1}_{\{s_k=i\}} = \sum_{k=1}^{m}\mathbf{1}_{\{s_k=j\}}, \\
	\mbox{then}\ \arg\min_{1\leq k\leq m}\mathbf{1}_{\{s_k=i\}}  < \arg\min_{1\leq k\leq m}\mathbf{1}_{\{s_k=j\}}
	\end{array}
	\right\},
	\end{align*}
    \endgroup
	 which is the reduced state space for $m$-history.
\end{definition}

Note that the first constraint regulates the number of the smaller index should not be less than the number of the larger index. And the second constraint regulates if two indices appear the same number of times, the smaller index should appear first.

After defining the standard state space and reduced state space, now we are going to construct a mapping between them. And this is trivial since we can directly get the mapping by the definition of reduced state space.

\begin{definition}[Reduced mapping]
	A reduced mapping from standard state space $S_s$ to reduced state space $S_r$ denoted as 
    $R:S_s\rightarrow S_r$. $\forall s = (s_1,...,s_m)\in S_s$, $R(s)$ is defined as follow
    \begingroup
	\begin{align*}
    R(s) = (s_1',...,s_m')\in S_r\ s.t.\ & \forall i\neq j,s_i'=s_j'\Leftrightarrow s_i=s_j,\ \\ 
    and\ & \forall i\ s.t.\ s_i=0\Leftrightarrow s_i'=0
	\end{align*}
    \endgroup
\end{definition}

Here, we give an example of a PDA PoW system.
For general application, one should observe the structure in their system and find a good way to reduce the number of states.

\subsubsection{Transition matrix}
In this step, we are going to formalize the transition function over the reduced state space, i.e., we apply the reduced mapping on the standard transition function. Suppose the standard transition function defined on standard probability space is $P_s:S_s\rightarrow S_s$. Then we define the reduced transition function as follows:
\begin{definition}[reduced transition function]
	Suppose $P_s:S_s\rightarrow S_s$ is the standard transition function, then the reduced transition function, denoted as $P_r:S_r\rightarrow S_r$, is defined as
	$$P_r(s_r,s_r') := \frac{\sum_{\{s:R(S)=Sr\}}\sum_{\{s':R(S')=Sr\}}P_s(s,s')}{ \mbox{number of }\{s:R(S)=Sr\} }$$
\end{definition}

\subsubsection{Stationary distribution}
To define the notion of stationary distribution in reduced state space, we need to specify the notion of the probability distribution over reduced stated space. We denote the space of probability distribution of reduced state space as
\begin{align*}
\mathcal{P}_r:=\{&p_r=(p_1,...,p_{|S_r|}): \\
&p_i\geq0\ \forall 1\leq i\leq|S_r|,\sum_{i=1}^{|S_r|}p_i=1 \}
\end{align*}
Note that we can view the stochastic process $W^{(1)},W^{(2)},...$ as another stochastic process of {\bf $m$-history}: 
\begin{align*}
& \{W^{(b)},W^{(b-1)},...,W^{(b-m+1)}\},\\
& \{W^{(b+1)},W^{(b)},...,W^{(b-m+2)}\},\\
& ...
\end{align*}
Moreover, we describe such stochastic process with random variables $W_m{(b)},W_m^{(b+1)},...$ where the support of $W_m^{(b)}$ is $S_r$. As a result, the distribution of $W_m^{(b)}$ can be represented by probability distribution in $\mathcal{P}_r$. That is, $W_m^{(b)}\sim p^{(b)}\in\mathcal{P}_r$.

Finally, we can define the stationary distribution of the stochastic process $W^{(b)}, W^{(b+1)}$, ... in the sense of $m$-history as follows:
\begin{definition}[reduced stationary distribution of $m$-history]
	Suppose $\bar{p}\in\mathcal{P}_r$, we say $\bar{p}$ is a reduced stationary distribution of reduced transition function $P_r$ if
	$$\bar{p} = P_r(\bar{p},\cdot)$$
\end{definition}

\subsection{Analysis}
Table \ref{table:result} shows the probability of consecutive winning with different system settings.
\begin{table}[h]
	\centering
	\caption{Probability of consecutive winning.}
	\resizebox{\columnwidth}{!}{\begin{tabular}{|l|l|l|l|l|l|l|}
        \hline
		n$\backslash$k& 1    & 2       & 3       & 4       & 5       & 6       \\ \hline
		1 & 1    & 1       & 1       & 1       & 1       & 1       \\ \hline
		2 & 0.5  & 0.19    & 5.56e-2 & 1.18e-2 & 1.80e-3 & 1.97e-4 \\ \hline
		3 & 0.34 & 7.30e-2 & 1.04e-2 & 9.41e-4 & 5.39e-5 & 1.94e-6 \\ \hline
		4 & 0.25 & 3.83e-2 & 3.52e-3 & 1.93e-4 & 6.25e-6 & 1.20e-7 \\ \hline
		5 & 0.20 & 2.35e-2 & 1.59e-3 & 6.14e-5 & 1.35e-6 & 1.70e-8 \\ \hline
		6 & 0.17 & 1.59e-2 & 8.46e-4 & 2.52e-5 & 4.17e-7 & 3.84e-9 \\ \hline
		7 & 0.14 & 1.45e-2 & 5.03e-4 & 1.21e-5 & 1.60e-7 & 1.14e-9 \\ \hline
	\end{tabular}}
	\label{table:result}
\end{table}
\vspace{-0.5em}

\section{Discussion}\label{sec:discussion}
We have shown that the probability of consecutive winning is drastically decreased after using the personalized difficulty adjustment proof-of-work mechanism. See Table~\ref{table:result}. The likeliness of double spending is decreased as the result. In this section, we will first summarize the results and compare with other's works. Then, discuss the major assumption, address identifiability, in this work. In the end, we will elaborate on some future works and open questions.

\subsection{Summary and comparison}

In Table~\ref{table:summary}, we summarize the consecutive winning probability from different mechanisms. The first row is the consecutive winning probability computed from the program in \cite{nakamoto2008bitcoin} and the following two rows are the results from the PDA for PoW mechanism we have proposed.

\begin{table}[h]
	\centering
	\caption{Summary: Attacker has 10\% computing power.}
    \resizebox{\columnwidth}{!}{\begin{tabular}{|l|l|l|l|}
    \hline
    Mechanism$\backslash$ k & 1      & 2      & 3      \\ \hline
    Bitcoin PoW               & 0.2046 & 0.0510 & 1.312e-2 \\ \hline
    PDA PoW: 2-exponential     & 0.1011 & 0.0054 & 1.560e-4 \\ \hline
    PDA PoW: 5-exponential     & 0.1030 & 0.0024 & 1.218e-5 \\ \hline
    Mechanism$\backslash$ k & 4        & 5         & 6       \\ \hline
    Bitcoin PoW               & 3.455e-3 & 9.137e-4  & 2.428e-4  \\ \hline
    PDA PoW: 2-exponential     & 1.214e-5 & 1.953e-8  & 8.447e-11 \\ \hline
    PDA PoW: 5-exponential     & 1.412e-8 & 3.660e-12 & 2.135e-16 \\ \hline
    \end{tabular}}
	\label{table:summary}
\end{table}
\vspace{-0.5em}


Clearly, the consecutive winning probability of PDA mechanism decreases much faster than that of traditional PoW setting. However, the double spending criteria in two mechanisms are actually not exactly the same. The traditional PoW mechanism allows forking in their blockchain, so the double spending probability in their estimation will be a little higher than that under the address-identifiability assumption. 

\subsection{Address identifiability}\label{sec:addressidentifiability}
In the paper by Satoshi Nakamoto \cite{nakamoto2008bitcoin}, he abandoned address identifiability in favor of complete anonymity. As a consequence, the traditional PoW system such as Bitcoin allows forking to happen in the blockchain and thus requires an analytical model that is different from our high-order Markov chain. In the analysis of our personalized difficulty setting, we adopt the address identifiability assumption so that the double spending probability will be smaller than that in traditional fork-style PoW system.

\section{Conclusion}\label{sec:conclusion}
In this paper, we solve the intrinsic overhead problem of proof-of-work-based blockchain by proposing a new PoW mechanism. We first formalize it as a voting system and then generalize it to a personalized difficulty adjustment system. The adjustment mechanism balances the winning distribution in the network because those who win a lot recently will less likely become the next verifier. Next, we use a high-order Markov chain to quantitatively model the PDA system. Finally, we show that the consecutive winning rate drastically decreases from 0.02\% to 0.00000008\% after adopting the exponential non-ordered difficulty function. As a result, PDA successfully decreases the probability of double spending by proposing a modified PoW protocol instead of the traditional approaches via information propagation. 

However, the performance of accelerating the transaction confirmation is not fully examined, which is a potential future work. On the other hand, we would like to point out the possible breakdown of a balanced motivation system. Since we increase the difficulty for those who have won recently, they might decide to have a break after winning and thus damage the dynamics of the system. Related problems were studied by Rosenfeld \cite{rosenfeld2014analysis} and Kroll et al. \cite{kroll2013economics} under the traditional setting. Another future work is the analysis of the winning distribution under PDA mechanism. It is possible that the winning distribution under PDA may not incentivize some participants, thus the PoW process suffers sabotage. We are actively working on the analysis and its implications.


\vspace{0.5em}
\begin{spacing}{0.75}
  \bibliographystyle{IEEEbib}
  \bibliography{refs}
\end{spacing}

\vspace{1.0em}

\appendices
\label{appendix}

\section{Proof of Theorem~\ref{thm:proportional}}\label{sec:proofthm1}
Computing power $C_i(b)$ is the rate to calculate a hash value; difficulty $D_i(b)$ here is the upper bound value the participant is required to solve.
For participant i, we model the waiting time of solving the hash value for blocks as an exponential random variable with mean $\alpha\times\frac{C_i(b)}{D_i(b)}$.
Given $C_i(b)$ and $D_i(b)$ with respect to a block $b$, the waiting time $T_i$ is 
\begingroup
\begin{align*}
T_i\stackrel{iid}{\sim}\exp(\alpha\frac{C_i(b)}{D_i(b)})=\exp(\alpha_i), where\ i=0, 1, ..., n-1
\end{align*}
\endgroup
$P_i^{(b)}$ is the probability that
\begingroup
\begin{align*}
P_i^{(b)}
& =P(T_i<T_1, ..., T_i<T_{i-1}, T_i<T_{i+1}, ..., T_i<T_n) \\
& =\int_{0}^{\inf}\prod_{j=0,j\neq{i}}^{n-1}P(T_j>x\ |\ T_i=x)\,P(T_i\in{dx}) \\
& =\int_{0}^{\inf}\prod_{j=0,j\neq{i}}^{n-1}e^{-\alpha_jx}\alpha_ie^{-\alpha_ix}\,dx \\
& =\alpha_i\int_{0}^{\inf}e^{-\sum_{j=0}^{n-1}\alpha_jx}\,dx \\
&= \frac{\alpha_i}{\sum_{j=0}^{n-1}\alpha_j}  = \frac{\alpha\frac{C_i(b)}{D_i(b)}}{\sum_{j=0}^{n-1}\alpha\frac{C_j(b)}{D_j(b)}}  \sim{\frac{C_i(b)}{D_i(b)}}
\end{align*}
\endgroup
Therefore, the probability $P_i^{(b)}$ is proportional to ${\frac{C_i(b)}{D_i(b)}}$.

\end{document}